\documentclass[conference]{IEEEtran}

\usepackage{cite}
\usepackage{amsmath,amssymb,amsfonts}
\usepackage{algorithmic}
\usepackage{graphicx}
\usepackage{textcomp}
\usepackage{xcolor}
\def\BibTeX{{\rm B\kern-.05em{\sc i\kern-.025em b}\kern-.08em
    T\kern-.1667em\lower.7ex\hbox{E}\kern-.125emX}}
\begin{document}

\title{Myths and Misconceptions about Attackers and Attacks}

\author{
\IEEEauthorblockN{Stjepan Groš}
\IEEEauthorblockA{\textit{Faculty of Electrical Engineering and Computing} \\
\textit{University of Zagreb}\\
Zagreb, Croatia \\
stjepan.gros@fer.hr}
}

\maketitle

\begin{abstract}
This paper is based on a three year project during which we studied attackers' behavior, reading military planning literature, and thinking on how would we do the same things they do, and what problems would we, as attackers, face. This research is still ongoing, but while participating in applications for other projects and talking to cyber security experts we constantly face the same issues, namely attackers' behavior is not well understood, and consequently, there are a number of misconceptions floating around that are simply not true, or are only partially true. This is actually expected as someone who casually follows news about incidents easily gets impression that attackers and attacks are everywhere and every one is under attack. Our goal in this paper is to debunk these myths, to show what attackers really can and can not, what dilemmas they face, what we don't know about attackers and attacks, etc. The conclusion is that, while attackers do have upper hand, they don't have absolute advantage, i.e. they also operate in an uncertain environment. Knowing this, means that defenses could be well established.
\end{abstract}


\section{Introduction}
\label{sec:introduction}

Cyber security experts have the primary task of protecting systems from attacks. Scientists, on the other hand, have the task of inventing a new methods of protection. Some do this at the system level, others at the component level - which itself could be a system when broken down into smaller components. Whatever and however they do it, i.e. protect these systems and components, it is often the case that they have discussed (or should discuss) what could happen \textit{before} they start protecting the system. In other words, it is important to know what it is against what we are protecting ourselves. In some areas there is a more or less good formal model that allows us to define and think about enemies, e.g. in cryptography \textit{random oracle model}, in design and development \textit{threat modeling} and in operations \textit{risk management}. However, if we exclude cryptography, in all the other cases there is no formal model of an adversary, and everything is based on the experience and knowledge of a person developing security measures. This dependence on knowledge and experience is not good, as it leaves a lot of room for errors. Moreover, because of the lack of the knowledge, we tend to have wrong assumptions, often based on myths and also on misconceptions.

Myths and misconceptions are dangerous because, on the basis of these myths and misconceptions, one tries to protect oneself from something that does not exist or has other characteristics than one assumes. So you end up defending from a different threat sources and threats than the ones that are really threatening you with potentially completely different capabilities, and that inevitably leads to security incidents. Even more often, however, there is a tendency to protect oneself from nothing in particular, i.e. one tries to protect oneself effectively from everything, and this is in vain, not to say almost impossible. This is particularly emphasized in research and R\&D projects when researchers start to fantasize about what they are going to do in their new brilliant research project  either because they really believe in it or because they believe that the reviewers will buy it from them, while at the same time they do not know what is really going on in the real world. What further underscores this is the abundant technical reports available on the Internet from many security companies, which are more concerned with the consequences of an attack than with the process, some of which is actually not known, but in any case they are not interesting for reports.

It is not that the knowledge necessary to break down these myths and misconceptions does not exist or is somehow kept secret. It is there, the problem is that this knowledge does not come from computer science, but from military science, intelligence and even geopolitics. So when computer scientists think of security problems, they tend to think too much in technical terms, not knowing that technical issues are only part of the problem.


Our aim in this paper is to expose these misconceptions and myths explicitly and to try to show why they are not, or at least not entirely, true. Although we have tried to argue each view, it is difficult or even impossible for some of them to prove that we are right (or wrong). So this may seem like a lot of hand waving, but our goal is actually not to convince you that we are right, but to get you to think about a broader picture of cyber security that goes beyond computer science and ventures into intelligence and military science. That is why this paper is aimed primarily at computer scientists and engineers, that is, those who are trying to develop new types of protection and detection tools.
 
Also, the reader should keep in mind that many myths and misconceptions are interrelated. We tried to structure them so that each myth or misconception is explained individually, but you'll realize that one myth as a consequence has another one, and if you deal with one, you'll also tackle the other. That said, we divided them and treat them separately in this text for the clarity.
 
The opinions expressed here were derived from the R\&D project \textit{Cyber Conflict Simulator} which has been running for almost three years. The goal of this project is to create a simulated environment in which incident handlers and management can practice what to do in the event of an incident. In this project, we explicitly try to avoid the technical level as it is not important for management. Since we were all engineers working on this project, this proved to be an excellent exercise to distance ourselves from the techniques and move more towards the management level. Also, in order to create something that the computer can calculate (simulate), we had to think intensively about attacks, how they happen and how the defenders become aware of the attack. Finally, we had to plan attacks in order to create exercises that further deepened our knowledge of the attacker's behavior.
 
We are primarily concerned with operational security, i.e. the security tools, techniques and procedures used on systems that are actively used by the intended users, i.e. during operation. The reason for this is that attackers primarily target these systems. The components of the system that are under development may be a target of the attack, but this also has to do with operational security - namely, organizations doing development have to establish operational procedures to protect their development processes and assets.

The paper is structured as follows. In Section \ref{sec:attackers} we analyze misconceptions and myths about the attackers themselves and then, in Section \ref{sec:attacks}, we concentrate on attacks. In section \ref{sec:discussion} we discuss would be changed in the approach to defense if we get rid of myths and misconceptions. The paper finishes with related work in Section \ref{sec:relatedwork} and conclusions in Section \ref{sec:conclusion} and a list of references.

\section{Attackers and their Behavior}
\label{sec:attackers}

While defenders and scientists develop mechanisms to increase the security of systems, attackers are generally treated as mythical, all-powerful creatures who can do whatever they want. Alternatively, they are treated as an undefined entity that represents every possible type of attacker. With such assumptions, discussions are very difficult, and the defence mechanisms developed are suboptimal, perhaps even harmful. In fact, during discussions, many different things are mixed up, and they are used as needed by anyone who wants to argue their case. To put this into perspective, let us take the software development process as an example. We all more or less know what happens to software that is developed based on requirements that are constantly changing. Here we have a similar case.

In the following subsections we list the assumptions and misconceptions often used in discussions that are either explicitly articulated, or more frequently, implicitly assumed.

\subsection{Misconception \#1: Attackers Do Simple Stuff}

This is actually an implicit rather than an explicit assumption. Namely, when discussing protective measures it is often assumed that the attacker will only take one tactical step and only this step needs to be prevented. For example, an attacker will perform a spear phishing attack, and we then focus on how to prevent a spear phishing attack, completely ignoring the fact that the attacker may use another means of initial compromise and that he will eventually succeed if he is sufficiently motivated. Worse, if this is our only defence, then after the initial intrusion the attacker can do whatever he wants.

Cyber kill chain \cite{hutchins2011}, as the most commonly used model of attacker behavior, even emphasizes this point. Of all the steps predicted for the attacker's behavior, the majority have to do with gaining a foothold in the target network. The last step, \textit{Actions on Objective}, actually brings together a lot of complexity that can be executed by attackers. This becomes even clearer when we consider that the average time between compromise and detection is several months \cite{muncaster2015}\cite{mandiant2015}. This means that an attacker has enough time to work on exactly this "simple" step \textit{Actions on Objective}.


The paper by Ahmad et. al. \cite{ahmad2019} also argues this. Namely they say that, based on their research, scientists as well as practitioners concentrate on technical level, mainly malware, while completely ignoring human component of attacks as well as decision-making processes that are behind the attacks.

Finally, in white paper for RSA Research \cite{kerner2015}, Kerner describes the process of reconnaissance which by itself requires more than just a simple stuff.

\subsection{Misconception \#2: Attackers Need to Find Just One Vulnerability}

This is a very common statement you may hear from people who work in the security field (e.g. \cite{tounsi2018}), and especially from those who want to convince you that you need to devote more resources to your security efforts. Actually, the full quote is \textit{defenders have to remove all the vulnerabilities to succeed, while the attackers have to find just one to succeed.} The earliest mention of this quote I could find dates back to 1990 \cite{national1990}, while it was probably popularized in 1998 by Bruce Schneier \cite{schneier1998}.

The statement itself is true, the attackers really do have the upper hand. But it's not that simple. First, the attackers need more than one vulnerability, that is, if they are not interested in web defacement and such simple attacks. Second, it's not as easy to find a vulnerability as it may sound at first. It is even more difficult if the potential victim follows a basic security hygiene. Finally, would phishing be the most common attack vector if so many vulnerabilities were waiting for an attacker?

\subsection{Myth \#1: Zero Day Vulnerabilities are all Around and Abundant}

Let's first distinguish between zero-day \textit{exploit}, and zero-day \textit{ vulnerability}. Zero-day vulnerability detection means that bugs in software are detected. This is very difficult, and there is a lot of activity to do it, but basically it is an unsolved problem. A formal verification could help, but there are certain assumptions as well as the problems with scalability that make this method not so accepted in the real-world, so it is of limited use and only works in special cases. The bottom line is that zero-day vulnerabilities is something that is/should be done in software development, not in operations. Therefore, we are not interested in zero-day vulnerability in this paper.

Zero-day exploits, on the other hand, are interesting from the operational point of view and when scientists discuss novel protections, or security engineers discuss protection mechanisms, they use the term in two different ways, ($i$) discussion on how to detect zero-day exploits, and ($ii$) when discussing attacker's capabilities.

Very frequently, when brainstorming on how to detect zero-day exploits, proposals about  \textit{monitoring a large number of networks} emerge. The idea is that it is possible to somehow detect zero-day exploit in that way. Yet, there are two fundamental problems with that idea. \textit{The first problem} is that you assume that you'll detect something unknown. In other words, how are you going to know that something is exploit, and more specifically, it is a zero day exploit when you never saw it? One viable approach to this problem is to turn unknown into known. Specifically, companies having products that detect attacks (e.g. intrusion detection systems) try to turn unknown into known by offering bounties to vulnerability researchers \cite{zdi}. Through those bounties they buy exploits as soon as they are discovered and build detections into their products. \textit{The second problem} is that zero days are expensive. For example, the company Zerodium published prices for different zero-day vulnerabilities (and exploits) on their Web pages \cite{zerodium}. Some vulnerabilities are in a range of several million dollar figures, so, when someone pays so much then they are very carefully where to use zero-day exploits because the moment it is used, it can be detected and patched. So, zero days are not widely used \cite{bilge2012}. Also, the analysis of APT attack shows that zero-days are not so frequently used as one might think \cite{ussath2016}\cite{li2016}\cite{joyce2016}\cite{ablon2017}.

The second assumption is that zero-day vulnerabilities \textit{give attacker great advantage against the victim}. This is to some degree true, i.e. having zero day allows attacker to perform some tactical step without victim being able to defend itself. But, it is also the fact that zero days exploits are not enough to pull of the whole operation. So, it doesn't mean that having zero-day vulnerability means automatically game over for defenders. It is a bit more complicated than that.

To illustrate the previous points, let's take as an example paper by Tounsi and Rais \cite{tounsi2018}. In their motivation section they claim (emphasis added by the authors of this paper):

\begin{quote}
    \textit{Those defenses, built for a previous generation of attacks, rely heavily on static malware signature-based or list-based pattern matching technology. This approach leaves those defenses extremely vulnerable to \textbf{ever evolving threats that exploit unknown and zero-day vulnerabilities.} What is therefore needed is a real-time system for information and intelligence sharing, in order to identify threat agents and targeted assets rather than to perpetuate the endless cycle of signature scanning.}
\end{quote}

First assumption in that paragraph is that zero-day vulnerabilities are big enough problem to motivate the whole paper. The second assumption is that it is possible somehow to solve the problem of zero-day exploits by information sharing. In order for that to be possible, two conditions have to be fulfilled. The first is that you can somehow detect the first zero-day attack, and the second is that there is at least one more victim targeted with the same attack. Both assumptions, as we saw, are far fetched.

\subsection{Myth \#2: Attackers can do, and do, whatever they want}
\label{subsec:dowhatever}

The consequences of this misconception are reflected in several ways during discussions. The first consequence is that people think that an attacker can be found anywhere within the network. What is completely ignored is that attacker has to somehow reach each point in the network, and that some points are more while others are less reachable. Not only that, but attackers try to be efficient, not to waste their time on unimportant parts of the victim's network. In essence what is missing is analysis of lateral movement, i.e. where the attacker can be, and if it is possible for the attacker to reach some point in the network.

The second consequence occurs in cyber security exercises in which scenarios are created without taking into account what attackers want and might do. Instead, some fictitious situations are invented. This is not the only problem with cyber security exercises, but it is certainly one of the more frequent ones.

This myth is also present in asset-based risk assessment. This is because asset-based risk assessment does not take into account how a threat can reach a particular asset, and so some error is introduced into the process by assuming that the attacker may be on an any asset (e.g. a database) when in reality this may not be the case. Opposite error is to ignore a path that the attacker can use to get to a particular object. This can happen because, e.g. assets that have no inherent value are usually ignored, while forgetting that they might be a stepping stone to a more valuable asset. In either case, we have the problem of not analyzing how an attacker can move through the network, or alternatively, that it is assumed he can move at will.




\subsection{Misconception \#3: Mixing Different Types of Attackers}

It's not the same if you are attacked by some individual, criminal group, or by an APT. That's so obvious that immediately almost everyone will jump and claim there's no misconception there. But, how many times did you discuss what an attacker can do, while at the same time not doing threat modeling before, or at least discussing who are the attackers that you are talking about?

For example, in Leszczyna and Wróbel \cite{leszczyna2019} they are describing threat intelligence platform. Yet, nowhere they say who's attacking and how. They have model of the adversary in which they say which adversaries they considered, but this model and its relation to threat intelligence is very confusing.


\subsection{Misconception \#4: Attackers just Hack Around}

When dealing with attackers, we tend to concentrate on specific tactical steps that attacker is performing, e.g. compromising some computer. From this, as a consequence, it follows that attackers have motives that guide them to do what they are doing. The contradiction is that it is almost well known that certain types of attackers have clear motives, e.g. motive for cyber criminals is profit. Yet, this motive isn't well connected with defense, i.e. when dealing with such threat actors, their motive is something not taken into account. The consequence of this thinking is that attackers are seen as just hacking randomly around, and hacking on whatever they can lay their hands down. Further consequence is that protections are then done so that everything is protected from this random hacking. 

Again, in white paper for RSA Research \cite{kerner2015}, Kerner describes the process of reconnaissance done by APT groups before engaging into attack. It is clear that the more thorough preparation of the attack, the less "just hack around" stuff.


\subsection{Misconception \#5: Confronting Attacker on the Technical Level}

All too often, when developing defenses, one tries to confront attacker only (or dominantly) on the technical level and prevent his tactical step. This is implicitly done by using a variety of technical approaches to network defense, e.g. introducing network intrusion detection systems (NIDS), SIEMs, building big data platforms of different kinds, using machine learning for anomaly detection, concentrating on fight against malware, or even by establishing SOCs.

While definitely important mechanisms for detection and prevention, by themselves they are not enough. First, attacker can change technical features very easily \cite{bianco2013} and easily circumvent defenses. Secondly, by concentrating on technical level we miss very important aspects of attacker's behavior.

The case in which big emphasis is placed on technical level is, for example, in ENISA document on good practices on interdependencies between OES and DSPs \cite{enisa2018}. In Table 1 they list threats, every single one being malware. But this list conflates a lot of different malware that have different modes of operation, and different modes of operation require different defense tactics. Petya and NotPetya are autonomous malware that have to be confronted on the technical level with basic security hygiene (in this case, updates), and detection. This works well here because NotPetya/Petya attack it is very fast and automated process, but also because IoCs for those malwares are very stable.

But other malware is different and fight on the technical level is very hard, not to say impossible. Stuxnet, for example, is also autonomous, but specially written for a specific environment, and thus highly adapted for this environment. Due to the lack of the connectivity of the target network to the Internet, it is also highly autonomous. On the technical level alone, it is next to impossible to defend from it. First, how to detect it? How to detect something you don't know exists? This question is the wholly grail of virus protection and intrusion detection. And if you manage to detect it, the attacker can adapt to your detection mechanisms. BlackEnergy brings this to the whole new level because it is a gateway for the attacker into a victim's network.

Here's an example. Let's say an attacker tries to infiltrate your organization and you detect phishing mail. First, it is hard to know its a spear phishing mail, but lets say defenders realize that at some point. Short of ignoring this case, tactical defense would be to train employees to avoid phishing. But if attackers are persistent they'll try until they succeed, so tactical defense is not enough. First very important information to know is that you are a victim of a targeted attack. 

\subsection{Misconception \#6: Defending against APTs using tactics applicable to Cyber Criminals}

More often than not I hear ideas about using strategy that might help in the case of defense from cyber criminals to be used to defend from APTs. The reasoning goes something like this, we'll monitor networks in many different networks and that way we'll be able to detect APT. The reason that this works in the case of cyber criminals is that they attack multiple targets with the same TTPs, and that way when the first victim detects the attack it can share IoCs with others and thus prevent attacks in other places. But, when we have APT there are two fundamental problems with this approach. The first problem is that APTs don't attack as many targets as cyber criminals, they are targeted. So, how are you going to detect an APT attack when it is attacking a single, or at most few, entities? The second problem is related to the fact that APTs have a lot of resources and can adjust to defenses \cite{sommer2010}.

\subsection{Myth \#3: Everything is clear for attackers}

This is assumption that attackers, when they penetrate target network, immediately know everything about that network. So they are very efficient and quickly reach their objectives. 

The truth is that attackers, after penetrating target/victim network, don't know much about the environment, so they have to do reconnaissance. During reconnaissance, they have to balance speed and stealthiness, i.e. the quicker they do reconnaissance the more likely is that they'll be detected, while the smaller chance of being detected means longer time to do reconnaissance. Intelligence preparation might help attackers in being quicker, but nevertheless without direct contact with target network, they are always left with uncertainties.

\section{Attacks}
\label{sec:attacks}

In this section we will analyze the attacker's side and show how neglecting this part can lead to a weak defense. Before we begin, let's define a few terms so that we avoid confusion because someone is using these terms for something else:

\begin{itemize}
    \item \textit{event} -- something that occurred in the network, malicious or benign.
    \item \textit{observable} -- consequence of an event that can be detected using different detection mechanisms.
    \item \textit{incident} -- observable that is a consequence of an attack.
    \item \textit{attack} -- continuous attacker's activity that has a specific goal, and that produces a series of incidents.
\end{itemize}

\subsection{Mixing Offensive and Defensive Operations}

It happens that there is confusion between offense (attack) and defense regarding strategy, operation, and tactics. For example, someone mixes defensive strategy with offensive operation and then again with defensive tactics and techniques. But, offense and defense are different and should not be mixed.

Operations in the case of a defense are activities that the defense performs continuously to keep its systems secure. For example, collecting logs and analyzing them for indicators of malicious activity-something that is now partially focused in Security Operations Center. Another example of running operation security is making sure defenses are up to date with the latest patches and updates. Note that these operations are not only used by victims, but also by attackers who also want to be secure. There are cases where attackers have had lapses in their OpSec that have made them more exposed to the public \cite{opsecfailures}.

On the other hand, in the case of an attacker, operation means activities that the attacker performs to breach the security of an organization or individual. The tactics used for both, and also the techniques, are quite different. And this can be easily seen by looking at the list of tactics used by the attacker, for example in MITRE ATT\& CK pattern \cite{mitre}. It is clear that these are not used for defense. Very importantly, penetration testing and similar activities should not be confused with defense activities. They help organizations to detect vulnerabilities, but in themselves they are offensive operations (i.e., attacks).

\subsection{Every incident is important}

There is a tendency to treat every observable as equally important, that is, it must be detected, evaluated, and if it is an incident then must be acted upon. In practice, of course, it is impossible to treat every incident thoroughly because there is always either a lack of resources or people tend to get sloppy, or both. So the result is that it doesn't get done, but whenever you read a text or hear a video that talks about how to defend yourself you'll hear that you need to do it. Even more so, when someone selling a defense product convinces you to buy that product, they usually start throwing around huge numbers like \textit{there are several million attacks on businesses every day.} Well, there probably are, but 99\% of them are most likely totally unimportant and could be solved by basic security hygiene.

Anyway, trying to address every possible incident is a completely pointless exercise. First of all, lack of resources and human sloppiness will be with us for a very, very long time. But more importantly, if you're dealing with so many incidents, you're not actually defending yourself, because by definition you're always one step behind the attacker. There is research that tries to do better by combining observables into attacks, but this is not easy and this research has yet to find ways of how to do that properly \cite{kovavcevic2020}.

\subsection{Let's Correlate all the Available Data}

This is related to the previous point, that is, in different discussions it can be frequently heard someone suggesting that we should build a big data platform and collect as much data as possible. Then, by some magic we'll be able to detect attackers as soon as they do something. As a nice example of this approach, though not the only one, is \cite{leszczyna2019}. In this paper authors describe threat intelligence and correlation platform.

The problem is that correlation is just a mathematical formula that doesn't intrinsically care what meaning the numbers you put into it have. Numbers in, numbers out. \textbf{Meaning}, on the other hand, is something that the person using correlation should think about. And rarely do the authors justify what the reason is for the correlation they are doing, other than perhaps the vague idea that \textit{we are trying to figure out what the attackers are doing}.

There's a whole research area that tries to infer from observables attacks, and it is hard \cite{kovavcevic2020}. In this research different models of attacker's behavior are studied, and key problem are dataset so the results are not so good. Now, what chances does have someone trying \textit{just to correlate data?}.

\subsection{When something is detected everything is clear}

Too many people are not aware how hard it is to actually know what happened during an incident, and this is exemplified in so many cases. For example, when an incident happened to Git repository hosted by PHP project \cite{phpincident}, it wasn't immediately clear how attack was done. Or, let's take even, conditionally speaking, simpler example, malware itself. When Stuxnet was detected, it took some time until it was realized, and only partially, what its capabilities are. The same is true for NotPetya \cite{notpetya}. First, it was thought that it's ransomware, only later to be clear it actually is not.

Or what happens if you learn of a breach from someone outside your organization that your data is exposed. Did you leak that data, was it an accident or an attack, or did a partner leak data? You don't know, and you won't know for some time until you do an investigation into what happened. And if you think you can spot a breach on your own, think twice. First, according to a report by Mandiant, in 53\% of cases, victims are notified by someone else. Also, the gap of several months between breach and detection \cite{mandiant2015} clearly shows the problems.

\section{Discussion and Implications}
\label{sec:discussion}

The proverb \textit{If you know the enemy and know yourself, you need not fear the result of a hundred battles. If you know yourself but not the enemy, for every victory gained you will also suffer a defeat. If you know neither the enemy nor yourself, you will succumb in every battle.} from Sun Tzu is quoted so often in various texts that it is almost a cliché. Yet Sun Tzu has deeply influenced military thinking for millennia, and today when the world's militaries prepare to defend themselves, they identify their enemies thoroughly and observe them closely - especially with the help of intelligence. Based on the information they gather, they carefully plan their tactics, operations, and strategies.

So, who in cyber security we tend to create defense for unknown entities? Why we don't take into account who our enemy is, what they do, and based on that, we plan our defenses -- including forecasting, detection, incident handling, and recovery processes? The authors think that the reason is rooted in the fact that cyber security used to be restricted to information and communication technology, but in the mean time it evolved into multidisciplinary topic which, to be tackled properly can not be based on technical level only.

In the following text we discuss how this approach influences defenses.

\subsection{Know Your Opponent}
\label{subsec:knowyouropponent}

First and foremost, try to find out who your enemy is. One very important aspect of each potential opponent is what it is trying to achieve. In other words, if you are pondering whether some threat actor could target you, the first discriminator is the question \textit{"Do we have what they are after?"}

If you don't have what it is after, then you don't have a problem, if you have, but you can remove it, then again you don't have a problem. Otherwise, you should try to see how your opponent might achieve its goal. To determine how the opponent might achieve its goal is actually more of an art than science. One approach would be to look at how they already achieved that before. This can be done by using databases like MITRE ATT\&CK \cite{mitre}. Note that those are only tactical steps and you should bare in mind that you should actually try to hunt for operations. Also, you have to monitor those threat groups as they are not static, they constantly change and adopt \cite{dennesen2016}.


\subsection{Defense Tactics}

Based on who's targeting you, you can also employ some general defense tactics. Here we'll list possible defense tactics for some general adversarial groups. Details will obviously vary.

\subsubsection{Individuals and autonomous malware}

Basic security hygiene is enough to handle individuals as they don't have enough resources to plan and conduct complex operations, nor they have expertise for planning and running complex operations. By \textit{basic security hygiene} we mean actually on best practices, i.e. having your systems patched, properly protected, networks properly segmented, unnecessary services removed, etc. 

In addition, if your environment is such that you have some special systems, i.e. the ones that are not so common, then you are even more safe from individuals. The reason is that to attack those systems, some level of knowledge is necessary which might be inaccessible to individuals, and/or they would not be interested in pursuing that knowledge. For example, if you have factory with PLCs, then this is technology that is not so common, i.e. PLCs are not general IT equipment so attacking them requires specialized knowledge and customized approach.

Note that there lurks \textit{security by obscurity} tactics, but this tactics might help for those adversaries.

We include \textbf{autonomous} malware into this category. When we say autonomous we mean the malware of fire-and-forget type, like NotPetya. Because machines do not yet exhibit complex behavior they cannot be innovative nor perform complex operations and thus the defense is the same as for individuals - basic security hygiene.

\subsubsection{Cyber Criminals}

Cyber criminals are financially motivated so in case you have something that can be monetized by them, then you might be attacked. The trick here is to know what cyber criminals are monetizing and how and to identify opportunities for them within you organization. This is important for defense, but also in case of incident response -- which we discuss later.

Cyber criminals attack multiple targets reusing the same tools and techniques, so collaborative defense in this case might be very efficient. By \textit{collaborative defense} we mean connecting and sharing information with all those that are as similar to you as possible. For example, if you are a small bank, then similar to you are other small banks. Note that similarity is measured by a number of parameters, like geographic region, environment, etc. For example, it is not the same if the small bank is in, for example, Croatia or in UK. Attackers attacking bank in Croatia would have problem with language barrier, the fact that Croatia is not in the Eurozone, and also by the simple fact that UK is richer than Croatia.

Very likely, only a handful of cyber criminal groups are innovative, while the majority are copycats that mirror tactics of innovative groups. Those innovative ones are more dangerous and require more resources to be tracked. Copycats, on the other hand, are easier to defend as they reuse TTPs which are already known. In which group you are, depends on who you are. Innovative cyber criminal groups are probably trying to achieve high gain, i.e. the ratio of invested resources and profit. So, if in your case can have high gain with respect to other potential targets, then you could be a target, otherwise, it is less likely.

\subsubsection{APTs}

These are obviously the hardest to handle. Tactics used in case of individuals and cyber criminals will hardly help in this case. The reason is that APTs have enough resources to adjust much of their TTPs, while in the same time they don't attack randomly around. So, collaborative security is much less effective in this cases. It is a fallacy to think that by monitoring large parts of network will help you.

But then, if you are a potential target of APTs, then you might and should get help from resourceful ally, which in majority - if not all - cases would be a government. 

\subsection{Incident Handling}

In case of an incident, if you are lucky, then you know exactly what happened, and thus, you are one big step closer to contain the incident and recover from it. But if you are not so lucky, then you don't know what happened, and in the worst case you only know that something happened and not much more than that. In such cases, when you don't have a complete information, \textbf{there has to be a strategy} on how to deal with this uncertainty.

To be clear, this is something that has already being tackled. Huang in 1999 \cite{huang1999} made an analogy of intrusion detection and response to military operations. In military operations, commanders need to have some hypothesis about the adversary’s operational and strategic goals to decide on a proper reaction to his actions discovered on the tactical level, such as e.g. troop movements.

The strategy how to handle unknown incident doesn't depend on question how, but \textbf{who}, i.e. who attacked you? The moment you manage to guess who attacked you, you know a lot more that you did before, i.e.:

\begin{itemize}
    \item you know why you were attacked (i.e. what you have that they are after),
    \item you know their modus operandi, i.e. what TTPs they use and using that knowledge you can hunt for them more efficiently.
\end{itemize}

Now, if you know what they are after then you should start from those resources and try to protect them first. Then you work your way to the point where they made initial breach and close that. If you try to start with the breach point - which is not easy as there are a number of ways that they could breach you and investigating each one takes time. In the mean time they are doing you a damage. 

If there are multiple resources they could be after, then the best way is to try to identify which one is breached. This can be found out by, e.g. some clues that attacker gave you or were found on the Internet. In case there are no clues, you have to do risk assessment (or use your exsiting RA) and then allocate available resources on places where the biggest damage could be inferred to you.

The next step is the implementation of the strategy, which depends on technical level.

\subsection{Risk Assessment}

Risk assessment can also benefit from knowing your opponent. It is currently taken into account via probability of threat being realized, but it could be better.

As currently done, risk assessment is primarily asset based. In other words, the usual methodology to assess risk is to list assets, threats and vulnerabilities, attach some probability to threat exercising vulnerability and damage it causes and combine them somehow and that's it.

The problem with such approach is the fact that it is hard to assess threat probability unless you take into account threat actor and threat source! If you identified your opponents, as proposed in subsection \ref{subsec:knowyouropponent} and taking into account TTPs of those opponents, then you basically identified probabilities.

But note that risk assessment could also be used during incident handling, if it is properly done. Namely, risk assessment actually evaluates likelihood of different operations that could be executed by attackers. So, if you during an incident identified who attacked you, in the risk assessment there should be identified a mean on how those attackers operate and you can use this information to try to hunt for traces to confirm that.

Risk assessment could also help to identify who attacked you because in risk assessment different scenarios of attacks are evaluated. Each scenario is hypothesis in case of an incident and your goal is to find which hypothesis is true, i.e., which attack was realized. 

Finally, risk assessment and hypotheses created during risk assessment could be used as a basis for threat hunting. This is similar to use of risk assessment during the incident, but in this case you are just checking for indicators based on anticipated scenarios.

\section{Related Work}
\label{sec:relatedwork}

This is certainly not the first paper that addresses issues about not taking into account attackers' behavior, and some of the papers were used to support certain arguments given in this paper. What is different is that this paper collects all of arguments in one place, adds few extra things and directly points out to the problem we have with defense.

The strongest argument in favor of this paper is TIBER framework \cite{tiber} from European Central Bank. In essence, what ECB tries to do via TIBER is to make red teaming better by informing red teams of attacker's behavior which can then be simulated in security tests performed by red teams. This directly supports a need to know attackers and to concentrate on their behavior, instead of trying to defend from everyone, and in the end from noone.


\section{Conclusions}
\label{sec:conclusion}

Long time ago (in technology terms) attackers just hacked around poking for deficiencies in defense and exploiting them for fun and profit. But, those time have long gone and in the mean time attackers evolved and proliferated, and they are still evolving. Yet, it is as if cyber security experts coming from computer science and IT companies stuck in those old times and are still trying to combat those attackers \textbf{only} using the same old methods at the technical level. Worse, different reports that are abudant on the Internet also concentrate on technical details creating illusion that that is the most important part and clouding other important information about attacker that are harder to spot.

What is important to have in mind is that assumptions that used to be true, are not any more and they are now just misconceptions and myths. This has a real consequences, as it makes defense grossly inefficient. Attackers, as they develop, directly or indirectly use knowledge from management, military science and intelligence. For example, cyber criminals are becoming innovative enterprises with long supply chains, and APTs are becoming more and more organized and their actions well planned and executed. Building defense means that those developments have to be taken into account, otherwise we are destined for failures.

Just to not be mistaken, there are advances that go in right direction, like MITRE ATT\&CK pattern. The problem is that too few people know about them and their purpose and for that reason we are stuck.

\bibliographystyle{IEEEtran}
\bibliography{bibliography}

\end{document}